\documentclass[preprint,showpacs,preprintnumbers,prb,fleqn,floatfix,superscriptaddress]{revtex4}

\usepackage{amssymb}
\usepackage{epsfig}
\usepackage{amsmath}

\begin{document}

\title{Slowing hot carrier relaxation in graphene using a magnetic field}

\author{P. \surname{Plochocka}}
\email{Paulina.Plochocka@grenoble.cnrs.fr} \affiliation{Laboratoire
National des Champs Magn\'etiques Intenses, Grenoble High Magnetic
Field Laboratory, CNRS, 25 avenue des Martyrs, 38042 Grenoble,
France}
\author{P. \surname{Kossacki}}
\affiliation{Laboratoire National des Champs Magn\'etiques Intenses,
Grenoble High Magnetic Field Laboratory, CNRS, 25 avenue des
Martyrs, 38042 Grenoble, France} \affiliation{Institute of
Experimental Physics, University of Warsaw, Poland}
\author{A. \surname{Golnik}}
\affiliation{Institute of Experimental Physics, University of
Warsaw, Poland}
\author{T. \surname{Kazimierczuk}}
\affiliation{Institute of Experimental Physics, University of
Warsaw, Poland}
\author{C. \surname{Berger }}
\affiliation{Georgia Institute of Technology, Atlanta, Georgia, USA}
\author{W.A. \surname{de Heer }}
\affiliation{Georgia Institute of Technology, Atlanta, Georgia, USA}
\author{M. \surname{Potemski}}
\affiliation{Laboratoire National des Champs Magn\'etiques Intenses,
Grenoble High Magnetic Field Laboratory, CNRS, 25 avenue des
Martyrs, 38042 Grenoble, France}

\date{\today }

\begin{abstract}
A degenerate pump--probe technique is used to investigate the non equilibrium carrier dynamics in multi--layer
graphene. Two distinctly different dynamics of the carrier relaxation are observed. A fast relaxation ($\sim 50$~fs) of
the carriers after the initial effect of phase space filling followed by a slower relaxation ($\sim 4$~ps) due to
thermalization. Both relaxation processes are less efficient when a magnetic field is applied at low temperatures which
is attributed to the suppression of the electron-electron Auger scattering due to the non equidistant Landau level
spacing of the Dirac fermions in graphene.
\end{abstract}

\maketitle

Carrier-–carrier scattering due to the Coulomb interaction is usually regarded as the dominant process which governs
the dynamics of hot carriers in solids at very short time scales.\cite{Rota93} This is true for conventional
(semiconductor) two-dimensional systems, even when their energy bands are quantized into discrete Landau levels by the
application of a magnetic field. This is because Auger-type scattering processes between equidistant Landau levels,
formed from bands with parabolic dispersions, are extremely efficient.\cite{Potemski91} Indeed, Auger scattering has
long been considered as the main obstacle for the fabrication of tunable far-infrared laser based on inter-Landau level
emission.\cite{Aoki86} For example, hot carriers generated by a strong electrical field give rise only to weak
cyclotron emission.\cite{Gornik79,Ikushima04} However, the application of a magnetic field should considerably
influence the electron-electron scattering process in strongly non parabolic electronic systems.

Graphene, a single monolayer of hexagonally arranged carbon atoms, is a two-dimensional system with quite unique
electronic properties mostly related to its peculiar band structure.\cite{Novoselv05,Zhang05,Novoselv06,Neto09} In
particular, carriers at the $K$ and $K'$ points of the Brillouin zone, where the conductance and valence bands touch,
have a linear dispersion relation and therefore behave as massless Dirac particles. A direct consequence of the linear
dispersion is the rather unusual Landau quantization, $E_n=\pm\tilde{c} \sqrt{2e\hbar B |n|}$, where $\tilde{c}
\simeq10^6$~$m/s$ is the effective speed of light (Fermi velocity) in graphene. Thus, in graphene the Landau levels
follow a square root dependence with the magnetic field and are not equally spaced. Electron-electron scattering (Auger
processes) should therefore be strongly suppressed in magnetic field creating favorable conditions for cyclotron
emission and population inversion.\cite{Morimoto08} A first step in this direction would be to demonstrate the
suppression of Auger processes for carriers in graphene in a magnetic field. Previous ultra--fast spectroscopic
investigations~\cite{Dawlaty08,Sun08,Newson09, George08} show that the electron-electron scattering occurs on the
timescale of a few tens of femtoseconds in the absence of an external magnetic field. The influence of the Landau
quantization on electron-electron scattering has so far not been investigated to the best of our knowledge.

In this paper we investigate the dynamics of the non-equilibrium carriers in graphene measured using a degenerate
pump-probe technique which directly probes the occupancy of states well above the Fermi level. Two characteristic
relaxation times are observed. A fast process ($\sim 50$~fs) which broadens the photo-created distribution and a slower
process ($\sim 4$~ps) due to thermalization. Applying an external magnetic field leads to a significantly longer
relaxation for both processes. This is attributed to a reduction in the electron-electron scattering due to energy
conservation \textit{e.g.} Auger scattering is blocked due to the non equidistant Landau level spacing in graphene. The
similar behavior suggests that electron-electron (Auger) scattering plays an important role in both the fast and slow
relaxation processes.

The experiment was performed in the pump-probe configuration using a Ti:Sapphire mode-locked laser. The repetition rate
was $80$~MHz, pulse width $\sim50$~fs, central wavelength around $800$~nm ($1.6$~eV), and spectral width $\simeq
35$~nm. Most of the measurements have been performed using a pulse energy of approximately $1.7$nJ corresponding to the
maximum pump pulse power of $140$~mW. The pump and probe pulses had the same wavelength and were co-linearly polarized.
Both beams were focused at the same place on the sample with a spot size of $\sim200 \mu$m and an angle between pump
and probe beams smaller than $10$ degrees. The two beams were individually chopped with different frequencies $f_{1}$
and $f_{2}$ close to $2$~kHz. The change in the intensity of the probe pulse (differential transmission), at different
delays between the pump and probe pulse, was measured using phase sensitive detection at the sum frequency
($f_{1}+f_{2}$). The magnetic field was applied in the Faraday configuration, perpendicular to the decoupled graphene
layers. The numbers of carriers created by the pump pulse can be estimated for the maximum power of the pump pulse.
Using the accepted value of $2.5\%$ for the absorption of a single graphene layer, the number of photo created carriers
per layer is of the order of $10^{8}$, corresponding to a carrier density of $\sim 3\times10^{11}$~cm$^{-2}$. The
investigated samples contain a high number of graphene layers (between 70 and 100) grown in vacuum in an induction
heating furnace by the thermal decomposition method, on a (4H) SiC substrate.\cite{Berger04,Berger06} Experimental data
confirm that the investigated layers exhibit the Dirac like electronic spectrum so that the system can be considered as
a multi-layer graphene sample.\cite{Berger06,Sadowski06,Sadowski07,Faugeras07}

\begin{figure}
\begin{center}
\includegraphics[width=1.0\linewidth]{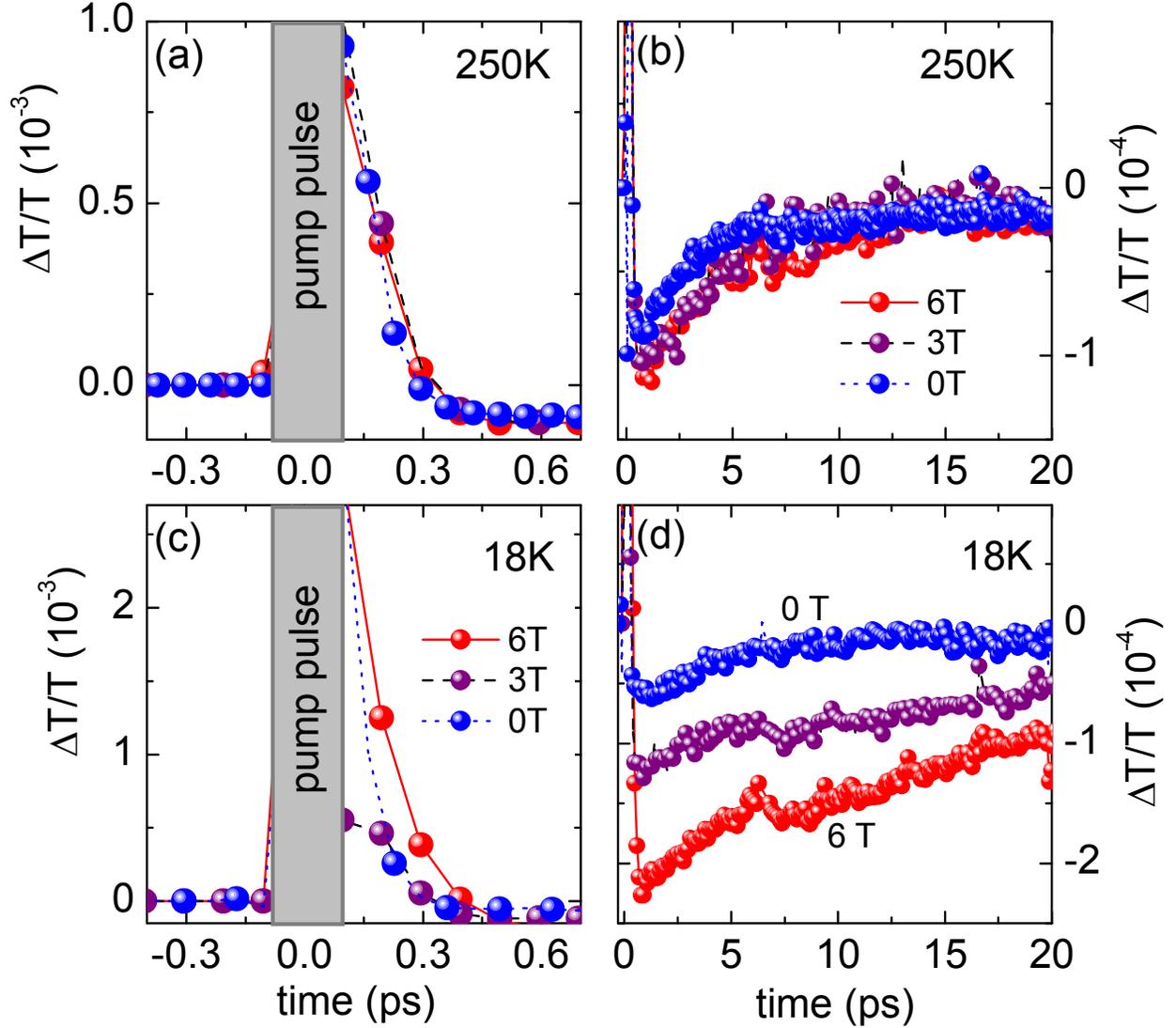}
\end{center}
\caption{(color on-line) The differential transmission $\Delta T/T$ as a function of the delay between the pump and
probe pulses measured at a temperature of (a-b) $250$~K and (c-d) $18$~K for magnetic fields in the range $0-6$~T.}
\label{fig1}
\end{figure}

The differential transmission $\Delta T/T$ as a function of delay between the pump and probe pulses, measured at
$T=250$K, is presented in Fig~\ref{fig1}(a-b) for magnetic fields in the range $0-6$~T. For negative delays (before the
pump pulse) the differential transmission $\Delta T/T$ is equal to zero. At zero delay (Fig~\ref{fig1}(a)), $\Delta
T/T$ increases significantly indicating that the sample becomes more transparent under the influence of the pump pulse.
Immediately after the pump pulse, a fast decrease of the signal is observed which is direct evidence that the
absorption increases. Interestingly, for delays longer than $\sim 300$~fs the differential transmission changes sign
becoming negative. Thus, under the influence of the pump pulse the absorption of the sample is larger that if there was
no perturbation at all. Subsequently, the negative signal decays with a characteristic time scale of the order of a few
picoseconds, much longer than the time scale of the initial decay (see Fig~\ref{fig1}(b)). It is clear from the data
that at high temperature ($250$~K) the magnetic field has almost no influence on the carrier relaxation.

\begin{figure}
\begin{center}
\includegraphics[width=0.9\linewidth]{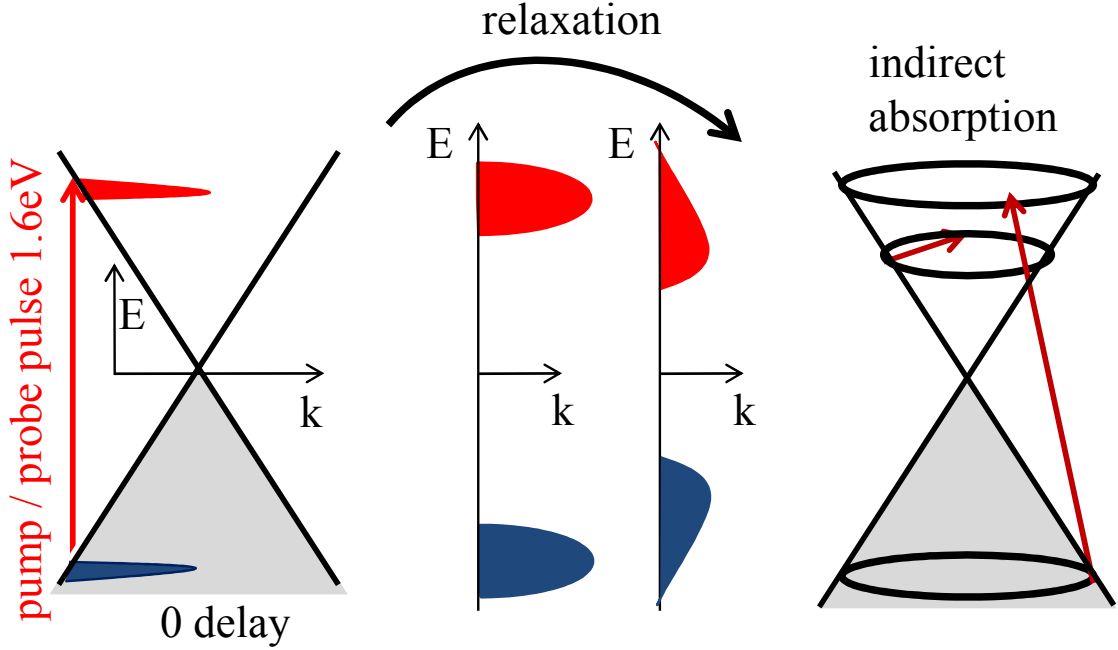}
\end{center}
\caption{(color on-line) Schematic illustration of the indirect absorption process (relaxed conservation of momentum)
for the electron-hole pair created by the probe pulse in the presence of a hot carrier distribution. The photo-created
distribution broadens and finally becomes asymmetric as the system cools. In the presence of the broad hot carrier
distribution indirect transitions involving the simultaneous scattering of an electron at a different energy are
allowed. As discussed in the text, relaxation involves both electron-electron and electron-phonon
scattering.}\label{fig2}
\end{figure}

The observed increase of the differential transmission under the influence of the pump pulse can be explained using a
simple phase space (Pauli) blocking argument.\cite{Dawlaty08, Sun08, Newson09, George08} The pump pulse creates a hot
electron-hole plasma with an approximately Gaussian distribution of carriers centered at $\pm 800$~meV and with a width
$\simeq 100$~meV. In our degenerate measurements the pump and probe pulses have the same wavelength so that the
absorption of the probe pulse at the same delay will decrease due to a blocking (filling) of the available states by
the carriers created by the pump pulse. At the energy of the pump pulse in the conduction band the number of available
states over the width of the pump pulse is of the order of $10^{13}$~cm$^{-2}$. This means that even at the maximum
pump pulse power used only a few percent of the available phase space at this energy is filled. Measurements at higher
power show that the increase in the differential transmission varies linearly with power~\cite{Dawlaty08} and the value
of $\Delta T/T$ measured in our experiment agrees with the data in
Refs.[\onlinecite{Dawlaty08,Sun08,Newson09,George08}] for similar excitation powers.

$\Delta T/T$ then decays quickly due to electron-electron scattering in which the photo-created carrier distribution
rapidly broadens as pairs of electrons scatter to lower and higher energy. The reduced phase space filling makes the
sample less and less opaque for the probe pulse. Our results are also in agreement with the findings of similar
measurements.\cite{Dawlaty08, George08} The characteristic decay time obtained by fitting an exponential decay to this
part of the data is of the order of $70$~fs, which is close to the temporal resolution of our experiment.

The subsequent increase in the absorption (negative $\Delta T/T$ ) can be attributed to indirect absorption involving
interaction with the now broad distribution of photo-created carriers (many body processes). It can be described as a
scattering which allows transitions with relaxed conservation of the momentum $\overrightarrow{k}$ of the electron-hole
pair created by the probe pulse, which is illustrated schematically in Fig.~\ref{fig2}. In contrast to the phase space
filling, which arises due to the population of photo-created electrons at the same energy as the pump excitation,
indirect absorption involves scattering with an electrons at different energies. The two processes compete, with phase
space filling dominating at very short delays. However, once the photo-created distribution broadens and decays in
energy, phase space filling is reduced, while many body processes involving scattering with electrons at different
energies remain possible. After a certain delay the carriers form a hot Fermi-Dirac distribution, however our
experiment is not sensitive to the particular shape of the carrier distribution. Experiments using a the probe pulse in
the the THz regime ~\cite{George08} suggest that a hot Fermi-Dirac distribution is formed around $150$~fs after the
pump pulse. One can think that at longer delays the many body processes increase the available phase space for
absorption at the pump/probe energy through non $\overrightarrow{k}$ conserving photon absorption. It is well
established that similar scattering processes at the Fermi level lead to the enhancement of the absorption and the
appearance of the so called Fermi edge singularity.\cite{Ruckenstein87} In our case, the probe pulse excites electrons
high in the Dirac cone well above the Fermi level. A similar enhanced absorption was recently observed in both
graphene~\cite{Li08} and graphite~\cite{Breusing09} and described in terms of a modification of the electronic levels
by the population of the photo created carriers. It seems likely that many different scattering processes are possible,
even those including more than two carriers. The scattering process presented in Fig.\ref{fig2} seems to be the most
probable due to the temporal evolution (cooling) of the distribution of the carriers, but it should be treated only as
a schematic representation.

Thus, it seems likely that the same scattering mechanism based on electron-electron interaction leads to the
enhancement of the absorption observed in our experimental data. The enhanced absorption decays with a characteristic
time of a few picosecond due to carrier thermalization via intra band carrier scattering (also Auger scattering) and by
inter band recombination of the photo-created carriers. In our experiment we are unable to distinguish between these
processes as we do not probe the occupation close to the Dirac point. Measurements with a THz wavelength probe pulse
[14] suggest that first there is a rather fast cooling of the broad distribution of carriers with inter band phonons or
cold carriers present in the sample on a time scale of 0.15-1ps leading to the thermalization of the carriers which is
followed by electron-hole recombination on the timescale of the order of a few picoseconds.

\begin{figure}
\begin{center}
\includegraphics[width=1.0\linewidth]{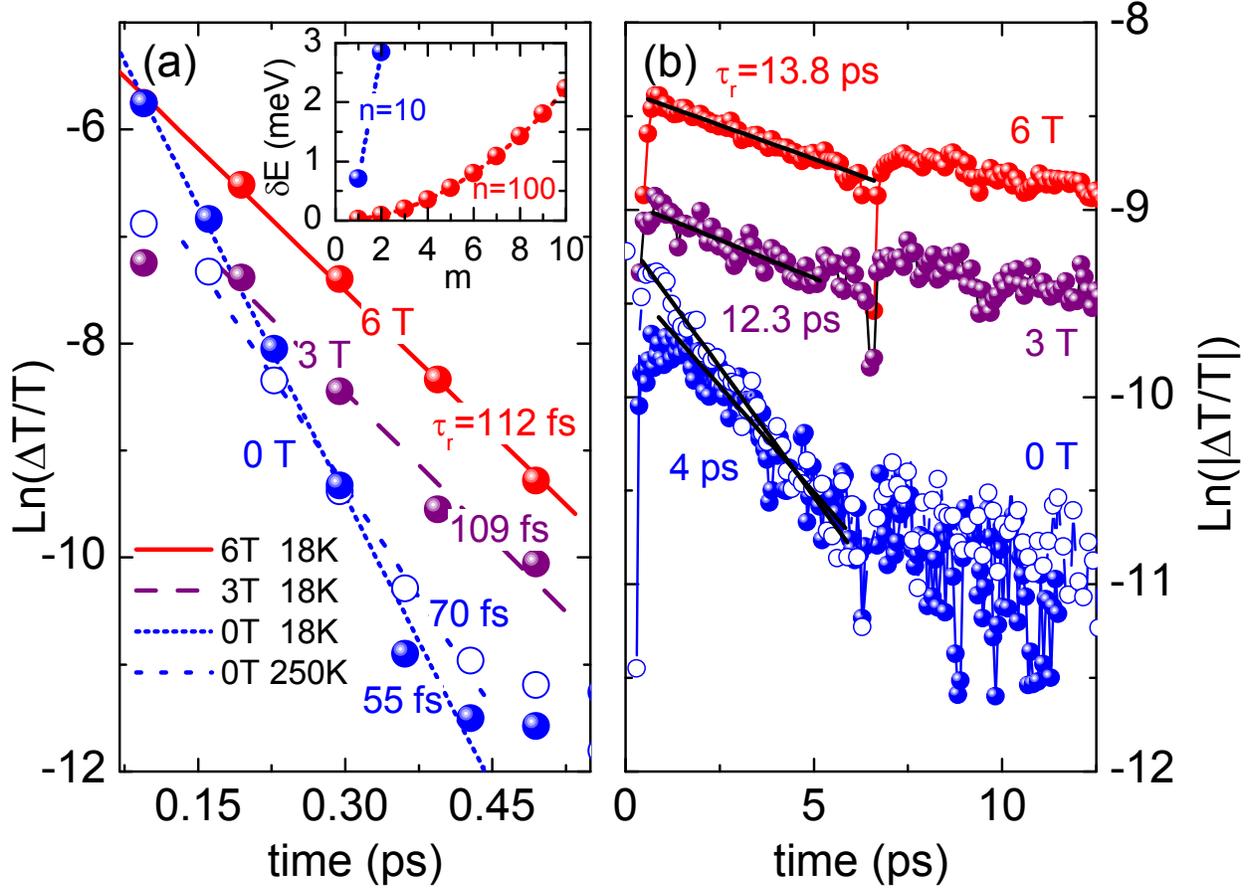}
\end{center}
\caption{(color on-line) (a-b) Natural log of the differential transmission as a function of a delay between pump and
probe pulses measured for magnetic fields ($0-6$~T) at $18$K. The $B=0$~T data measured at $250$~K is plotted using
open circles. The relaxation times $\tau_{r}$ extracted from linear fits are indicated. The inset of (a) shows the
calculated energy mismatch (Eq.~\ref{Eq1}) for Auger processes for carriers in the $n=10$ and $n=100$ Landau level
versus change in Landau level index $m$} \label{fig3}
\end{figure}

In Fig.\ref{fig1}(c-d) we plot $\Delta T/T$ versus delay time measured at a temperature of $18$~K for magnetic fields
in the range 0--6~T. The magnetic field has a considerable influence on both the fast relaxation (Fig.\ref{fig1}(c))
and the slow relaxation (Fig.\ref{fig1}(d)). For the latter, magnetic field greatly increases the amplitude of the
negative differential transmission, while at the same time slowing down the thermalization. This can be seen more
clearly in Fig~\ref{fig3} which plots $\ln(\Delta T/T)$ versus delay time to highlight the exponential character of the
relaxation. The characteristic time  of the decay ($\tau_{r}$) can be extracted from the slope of such a plot. For the
slow relaxation (Fig~\ref{fig3}(b)) at zero magnetic field there is little difference between the high and low
temperature data with a relaxation time of $\tau_{r}\sim4$~ps at both $18$~K and $250$~K. The relaxation becomes slower
by a factor of 3--4, changing from $\tau_{r}\sim4$~ps at $B=0$~T to $\tau_{r}\sim$ 12--14~ps for magnetic fields above
$~3$~T. The fast decay of $\ln(\Delta T/T)$ at low temperatures is plotted in Fig~\ref{fig3}(a) for different magnetic
fields. Data measured at zero magnetic field and $250$~K is also shown for comparison. Without magnetic field there is
little difference between the high and low temperature data with a relaxation time of $\tau_{r}\sim55$~fs at $18$~K and
$\tau_{r}\sim70$~fs at $250$~K. However, at low temperatures, the application of a modest magnetic field $\sim$ 3--6~T
doubles the relaxation time to $\tau_{r}\sim 110$~fs, providing direct experimental evidence that the electron-electron
scattering is significantly less effective in the presence of Landau quantization. This slow down, which is common for
both the slow and fast relaxation processes, can be seen as a proof that in both cases the electron--electron
scattering or thermalization of the hot plasma with the cold electrons is reduced in the presence of Landau
quantization which limits the possible energy of the initial and final states.

This is in contrast to standard 2DEG systems in which Auger scattering is enhanced \cite{Potemski91,Shahbazyan04} when
the zero magnetic field density of states collapses onto equally spaced Landau levels, increasing both the initial and
final density of states. In such an Auger process two electrons in Landau level $n$, scatter respectively to the $n+m$
and $n-m$ Landau levels ($m=1,2,\cdots$) conserving total energy. Processes involving the scattering of two electrons
from Landau level $n$, to Landau levels $n+m$ and $n-q$ with $m\neq q$, are forbidden due to quasi angular momentum
($k$-vector) conservation. It has been shown theoretically~\cite{Tsitsichvili97} that the most efficient scattering is
to the adjacent Landau levels ($m=1$). The total probability of scattering to the outlying Landau levels
($m=2,3,4,\cdots$) is roughly the same as the probability of scattering to the nearest Landau level ($m=1$). The
authors also show that the higher the Landau level index the larger the probability of the scattering with a change of
the index $m>1$. In principle, the non equidistant energetic separation of the Landau levels in graphene is expected to
suppresses Auger scattering since energy conservation is not fulfilled. An enhancement of the cyclotron emission has
been reported in p-type Germanium due to non equidistant light-hole Landau levels separation, however the situation is
complicated due to the light hole-heavy hole mixing.\cite{Kuroda91} In graphene, the energy mismatch for Auger
processes between Landau levels is
\begin{equation}\label{Eq1}
\delta E=\tilde{c} \sqrt{2e\hbar B}(2\sqrt{n}-\sqrt{n+m}-\sqrt{n-m}),
\end{equation}
with $n\approx 100$ for the absorption of $1.6$~eV radiation at $B=6$~T. The energy mismatch versus $m$ is plotted in
the inset of Fig.\ref{fig3}(a). The Landau level spacing is a few meV at this energy and this magnetic field,
comparable to the Landau level broadening~\cite{Orlita08} so that the density of states is expected to be modulated
even though discrete Landau levels are almost certainly not resolved. Under such conditions Auger scattering should not
be significantly suppressed at least for $m=1$. However, with increasing $m$, the energy difference increases reducing
the overlap in phase space for Auger scattering. Thus, under the experimental conditions employed we observe a
reduction of the scattering time of only around a factor of $2$. Exciting at lower energies (lower n), where the
separation between Landau levels is larger, should lead to a considerably more efficient blocking of Auger scattering,
as can be seen in Fig.\ref{fig3}(a) where we also plot the expected energy mismatch for $n=10$. Further pump probe
measurements at THz frequencies are planned to verify this hypothesis. The reduction in Auger scattering under magnetic
field reported here is an important and new observation since it potentially opens the way for cyclotron emission or
even lasing in graphene.

\begin{figure}
\begin{center}
\includegraphics[width=1.0\linewidth]{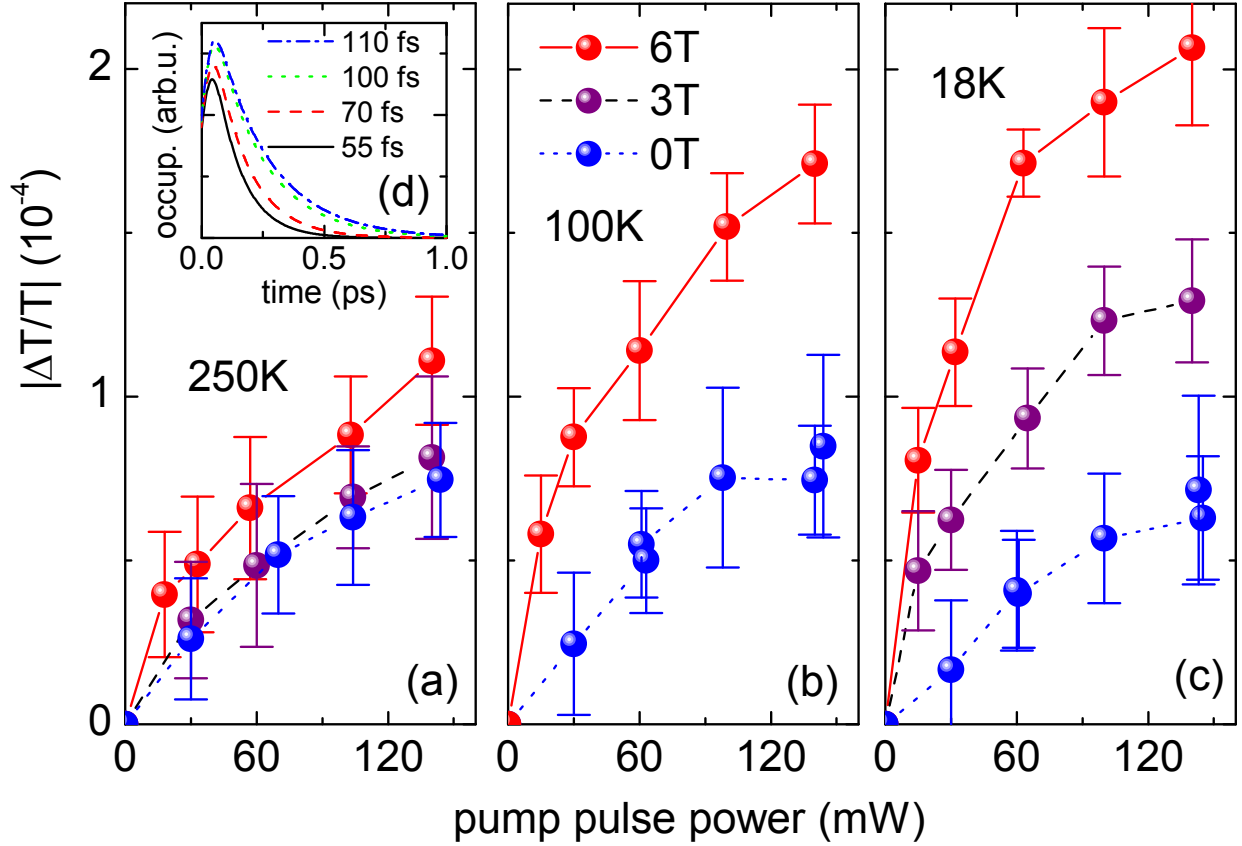}
\end{center}
\caption{(color on-line) (a-c) Pump probe power dependence of $|\Delta T/T|$ for a positive delay corresponding to the
maximum enhanced absorption (negative $\Delta T/T$) measured at different temperatures for magnetic fields ($0-6$~T).
(d) Results of a simple rate equation model; the calculated occupation of the state assuming different life times
(relaxation rates) and a Gaussian temporal shape for the generation function.} \label{fig4}
\end{figure}

Finally, to verify the hypothesis that the enhanced absorption (negative $\Delta T/T$) is related with scattering with
photo-created electrons we have investigated this effect as a function of the power of the pump pulse, temperature and
magnetic field. In Fig~\ref{fig4} we plot the differential transmission at a positive delay corresponding to the
maximum amplitude of the negative $\Delta T/T$ (maximum enhanced absorption). For the $250$~K data, presented in
Fig~\ref{fig4}(a), $|\Delta T/T|$ increases as the power of the pump pulse increases consistent with our interpretation
of an enhanced absorption due to indirect (non $\overrightarrow{k}$ conserving) processes involving scattering with the
carriers created by the pump pulse. As the number of the photo-created carriers increases with the power of the pump
pulse it is natural that the indirect absorption also increases.

At $250$~K an applied magnetic field has little influence since the data taken at different magnetic fields are almost
identical (as expected from the data presented in Fig~\ref{fig1} which were taken at the maximum power). $|\Delta T/T|$
measured at lower temperatures can be seen in Fig.\ref{fig4}(b-c). It is clear that with decreasing temperature the
magnetic field has an increased influence on the pump pulse power dependence of $|\Delta T/T|$. For the lowest
temperature of $18$K the signal at B$=6$T is almost as twice as large as at $0$T. The change in the amplitude of the
negative $|\Delta T/T|$ with a magnetic field can be explained using a simple rate equation model where we consider
pumping a level for 50fs (duration of the pulse) and then escaping from this level with a 70fs escape time (0T decay
time) and 110 fs (6T decay time) , which is presented in Fig~\ref{fig4}(d).  As the negative value of $|\Delta T/T|$ is
related to the number of carriers at an energy near the energy of the probe pulse it is strongly dependent on their
escape time. The role of temperature can be understood as follows; with increasing lattice temperature the elastic
electron-phonon scattering broadens the Landau levels and therefore suppresses the quenching of the Auger scattering.
One should note that this is true even if the carrier temperature is much higher than the lattice temperature. For
example, at a temperature of $50$~K the thermal energy ($k_B T$) is comparable to the Landau level separation.

In conclusion, a pump-probe technique has been used to probe the carrier dynamics in multi-layer graphene.  We find
that the hot carrier relaxation is slowed in the presence of a magnetic field. This is interpreted as a reduction of
electron-electron (Auger) scattering due to the unusual Landau quantization of Dirac fermions in graphene as well as a
proof of the importance of the electron-electron scattering compared to the electron-phonon scattering in all the
phases of the carrier relaxation and thermalization. Our measurements, which probe Landau levels with a high index
($n\approx 100$), suggest that for lower Landau levels, Auger processes may be completely suppressed. This makes
graphene a promising system for the implementation of the long ago proposed~\cite{Aoki86} tunable far infrared Landau
level laser.

\begin{acknowledgments}
This work has been supported by contract ANR-06-NANO-019, RTRA DISPOGRAPH FCSN-2008-09P, and CNRS PICS-4340. Two of us
(P.P. and P.K.) are financially supported by the EU under FP7, contract no. 221249 `SESAM' and contract no. 221515
`MOCNA' respectively.

\end{acknowledgments}


\end{document}